\PassOptionsToPackage{pdfpagelabels=false}{hyperref}
\documentclass[useAMS,usenatbib]{mnras}
\pdfoutput=1
\usepackage[pdftex]{graphicx}
\usepackage{amsmath}
\usepackage{amssymb}
\usepackage{natbib}
\usepackage{color}
\usepackage{times}
\usepackage{comment}
\usepackage{tabularx}

\def \BE{\begin{equation}}
\def \EE{\end{equation}}	
\def \BC{\begin{center}}
\def \EC{\end{center}}
\def \BEA{\begin{eqnarray}}
\def \EEA{\end{eqnarray}}

\def \eRO{\it eROSITA}
\def \eRA{\it eRASS}

\def \HI{ h^{-1}}
\def \MSUN{\mathrm{M}_{\odot}}

\def \FNL{f_{\rm NL}}

\def \SIGMA8{\sigma_{8}}
\def \OM{\Omega_{\rm m}}
\def \OB{\Omega_{\rm b}}
\def \OL{\Omega_{\rm DE}}

\def \M500c{M_{\rm 500c}}

\def \ALPHALM{\alpha_{\rm LM}}
\def \BETALM{\beta_{\rm LM}}
\def \GAMMALM{\gamma_{\rm LM}}
\def \SIGMALM{\sigma_{\rm{LM}}}

\def \etal{{\it et al.\ }}

\title[Dark-energy forecasts with $\eRO$]{Forecasts on dark energy from the X-ray cluster survey with $\eRO$: constraints from counts and clustering}

 \author[Pillepich et al.]
 {Annalisa Pillepich$^1$$^,$$^2$$^,$$^3$\thanks{E-mail: pillepich@mpia-hd.mpg.de},
 Thomas H. Reiprich$^4$,
 Cristiano Porciani$^4$, 
 Katharina Borm$^4$, \newauthor
 and Andrea Merloni$^5$
\\
$^1${Max-{\it Planck}-Institut f{\"u}r Astronomie, K{\"o}nigstuhl 17, 69117 Heidelberg, Germany}\\
$^2${Harvard--Smithsonian Center for Astrophysics, 60 Garden Street, Cambridge, MA 02138}\\
$^3${Department of Astronomy and Astrophysics, University of California, Santa Cruz, CA 95064, USA}\\
$^4${Argelander-Institut f\"ur Astronomie, Auf dem H\"ugel 71, D-53121 Bonn,Germany}\\
$^5${Max-{\it Planck}-Institut f{\"u}r Extraterrestriche Physik, Postfach 1312, 85741, Garching bei M{\"u}nchen, Gemany}
}

\begin{document} 
\maketitle

\begin{abstract}
We forecast the potential of the forthcoming X-ray galaxy-cluster survey with $\eRO$ to constrain dark-energy models. We focus on spatially-flat cosmological scenarios with either constant or time-dependent dark-energy equation-of-state parameters. Fisher information is extracted from the number density and spatial clustering of a photon-count-limited sample of clusters of galaxies up to $z\sim2$. 
We consider different scenarios for the availability of (i) X-ray follow-up observations, (ii) photometric and spectroscopic redshifts, and (iii) accurate knowledge of the observable -- mass relation down to the scale of galaxy groups.
With about 125,000 clusters (detected with more than 50 photons and with mass $\M500c \gtrsim 10^{13} h^{-1}\MSUN$) from an average all-sky exposure of 1.6 ks, $\eRO$ will give marginalized, one-dimensional, 1$\sigma$ errors of $\Delta \SIGMA8 = \pm~0.008$ ($\sim$1 per cent), $\Delta \OM = \pm~0.006$ (2.2 per cent), $\Delta w_0 = \pm~0.07$ (7 per cent), and $\Delta w_a = \pm~0.25$ (optimistic scenario) in combination with (and largely improving upon) current constraints from various cosmological probes (cosmic microwave background, BAOs, Type Ia SNe).
Our findings correspond to a dark-energy figure of merit in the range of $116-162$ (after the four years of all-sky survey), making $\eRO$ one of the first Stage IV experiments to come on line according to the classification of the Dark Energy Task Force. 
To secure improved mass calibrations and to include high-redshift clusters ($z \gtrsim 0.5$) as well as objects at the group-mass scale ($\M500c \lesssim 5\times 10^{13} h^{-1}\MSUN$) will be vital to reach such accuracies.
\end{abstract}

\begin{keywords}
cosmology: cosmological parameters, large-scale structure, early Universe -- galaxies: clusters -- X-rays: galaxies: clusters.
\end{keywords}


\section{Introduction}

Dark energy (DE) is invoked to explain the late-time accelerated expansion of the Universe \citep{Riess:1998, Perlmutter:1999}. In the last decade, ambitious observational programs have been designed with the ultimate goal of determining its physical properties.
Experiments are generally used to constrain a number of phenomenological
parameters that characterize dark energy and differentiate it from the cosmological constant. 
This is done notwithstanding the caveat that many DE parameterizations, however, cannot easily be associated with physical dark-energy models \citep{Scherrer:2015}.
Yet, different parameterizations of DE leave different signatures in the time evolution of the expansion rate, the growth rate of structure, and the angular-diameter distance vs. redshift relation \citep[e.g.][]{Albrecht:2006}, which, in turn, can be inferred via the measurement of galaxies and galaxy clusters statistics \citep[see e.g. the reviews by][]{Frieman:2008rev, Allen:2011rev, Huterer:2017rev}.

Samples of massive galaxy clusters for cosmological studies have been assembled in the last decade across wavelengths and redshift ranges, e.g. in the X-rays (e.g. REFLEX, NORAS, HIFLUGCS, 400d, MACS, XCS, XXL, X-CLASS), in the optical bands (e.g. maxBCG, RCS2, DES, PanSTARSS) and at submm wavelengths (e.g. with SPT, ACT, {\it Planck}), bearing counts from just a handful to a few thousands. Cosmological constraints on DE have been obtained from the evolution of their number density across redshift \citep[aka cluster counts;][]{Henry:2009, Vikhlinin:2009b, Mantz:2010, Rozo:2010, Tinker:2012, Benson:2013, Mantz:2015, deHaan:2016}, from their spatial clustering properties \citep{Schuecker:2003a, Schuecker:2003b}, from the measurement of their gas fractions \mbox{\citep[e.g.][]{Allen:2008, Mantz:2014}}, or from a combination of those \citep[e.g. more recently][]{Schellenberger:2017}.

In this paper, we use the Fisher information matrix to estimate
the constraints that the forthcoming X-ray all-sky galaxy-cluster survey with $\eRO$ \citep{Predehl:2006, Predehl:2010, Merloni:2012} will set
on the properties of DE together with other parameters of the standard cosmological model. $\eRO$ is the Extended ROentgen Survey with an Imaging Telescope Array (\url{http://www. mpe.mpg.de/erosita/}): it has been assembled and tested by a consortium of German institutions, will operate in the soft and hard X-ray energy bands (0.3-10 keV), and will be shipped to an L2 orbit on-board the Russian `Spectrum-Roentgen-Gamma' (SRG) satellite, planned (at the time of writing) to be launched from Baikonur in March-April 2019.
Equipped with a system of 7 identical co-aligned mirror modules, $\eRO$ will have a spatial resolution comparable to the XMM-Newton one, a similar effective area at low energies, but a wider field of view, while it will be 20-30 times more sensitive than the ROSAT sky survey in the soft band and the first all sky survey in the hard band. $\eRO$ is planned to perform a 4-yr-long all-sky survey and is expected to deliver a sample of about 3 million Active Galactic Nuclei (AGNs) and of about hundred thousands galaxy clusters up to $z \gtrsim 1$ \citep{Merloni:2012, Pillepich:2012}. 

The aim of this paper is to provide the community with baseline expectations for the $\eRO$ constraints on DE, by focusing on the abundance and spatial clustering of its photon-count limited, all-sky, galaxy-cluster sample. Our calculations are therefore meticulously tailored around the characteristics of the $\eRO$ telescope and the actual way observations will be taken. Here we expand on the methods and results of \cite{Pillepich:2012} where special attention was devoted to the possibility of detecting primordial non-Gaussianity and to simultaneously constraining cosmological and intra-cluster medium (ICM) parameters (the so called, self-calibration technique). 
In what follows, we quantify the benefits of (a) including X-ray follow-up observations to tighten the available priors on the adopted observable-mass relation (luminosity--mass); (b) augmenting availability and accuracy of redshift estimates for the $\eRO$-detected clusters (photometric vs spectroscopic redshifts); and (c) extending the sample to lower-mass objects yet at a fixed detection limit of 50 photons per cluster. 

The paper is organized as follows. In Section \ref{SEC:SCENARIOS},
we introduce the cosmological models used in our study, while, in Section \ref{SEC:SETUP}, we summarize our analysis setup. After giving some basic definitions, here, we present the statistical tools and observables upon which our forecast is based and also examine the $\eRO$ survey strategies.
The optimal choice of the redshift and photon-count bins and the necessity of follow-up observations to improve the knowledge on the X-ray scaling-relation parameters are briefly discussed in Sections \ref{SEC:BINNING} and \ref{SEC:SELFCAL}. 
The main results of the paper are given in Section \ref{SEC:RESULTS}, where we present the constraints which will be obtained with the $\eRO$ all-sky cluster survey after four years of observations (by itself and in combination with other leading cosmological probes). We also investigate the impact of high-redshift clusters and provide `early' $\eRO$ constraints that will be available after one year of observations as well as those derived from half-sky coverage only. 
We critically discuss our methodology and assumptions in Section \ref{SEC:MODELS}, where we also compare some of our Fisher forecasts with a full likelihood study. We conclude in Section \ref{SEC:CONCLUSIONS}.


%
\begin{table*}
\footnotesize
\begin{center}
\caption{Model parameters adopted in the analysis: fiducial values and priors used throughout unless otherwise stated. Values for the cosmology sector and X-ray luminosity--mass and temperature--mass relations are taken, respectively, from \textcolor{blue}{Komatsu et al. (2009)} and \textcolor{blue}{Vikhlinin et al. (2009a)}. $\mathcal N ( \sigma ) $ represents the normal distribution with mean given by the fiducial value and variance $\sigma^2$; $\mathcal U(x_1,x_2)$ denotes the uniform distribution with endpoints $x_1$ and $x_2$; and $\mathcal \delta_D[x] $ signifies that the parameter is kept fixed at the fiducial value $x$. Throughout the paper, we assume a flat cosmology, with $\OL = 1 - \OM$; Gaussian initial conditions, $\FNL \equiv 0$; and temperature--mass relation parameters by \textcolor{blue}{Vikhlinin et al. (2009)}, coupled to the luminosity--mass relation via a bivariate lognormal distribution with  null correlation coefficient between luminosity and temperature (see \textcolor{blue}{Pillepich et al. (2012)} for a discussion). Standard priors are taken from \textcolor{blue}{Riess et al. (2011)} for the Hubble Parameter ($\Delta h = \pm 0.022$), \textcolor{blue}{Cooke et al. (2014a)} for the mean baryon density from primordial deuterium measurements coupled to standard models of Big Bang nucleosynthesis ($\Delta (\OB h^2) = \pm0.0045$), and from \textcolor{blue}{Vikhlinin et al. (2009a)} for the luminosity--mass parameters.}
\label{TAB:PARAM}
\begin{tabular}{ l l c l}
\hline
 & & & \\
& Description 
& Fiducial Value 
& Adopted Priors \\
 & & & \\
\hline
 & & & \\
Cosmology  & & & \\
 & & & \\

$\SIGMA8$ 		& Normalization of $P(k)$						& 0.817 								& $\mathcal U(-\infty,+\infty)$\\
$\OM$ 			& Total Matter Fraction						& 0.279 								& $\mathcal U(-\infty,+\infty)$\\
$n_s$ 			& Spectral index 							& 0.96 								& $\mathcal U(-\infty,+\infty)$\\
$h$ 				& Hubble Constant 							& 0.701 								& $\mathcal N (0.022) $ \\
$ \OB$ 			& Baryon Fraction							& 0.0462								& see \cite{Cooke:2014a} \\
$ w_0$ 			& DE EoS Parameter: present value		& $-1$								& $\mathcal U(-\infty,+\infty)$ \\
$ w_a$ 			& DE Eos Parameter: rate of change 		& 0									& $\mathcal U(-\infty,+\infty)$\\

  & & & \\
X-ray Scaling-Relations  & & & \\
 & & & \\

$\ALPHALM$ 		& LM relation: Slope 								& 1.61 					& $\mathcal N (0.14) $ \\
$\GAMMALM$ 		& LM relation: $z$-dependent Factor 				& 1.85 					&  $\mathcal N (0.42) $ \\
$\BETALM $		& LM relation: Normalization 						& 101.483 				& $\mathcal N (0.085) $ \\
$\SIGMALM$	 	& LM relation: Logarithmic Scatter 					& 0.396 					&$\mathcal N (0.039) $  \\ 

& & & \\

$\alpha_{\rm TM}$ 	& TM relation: Slope 								& 0.65 						& $\mathcal \delta_D (0.65) $ \\
$\beta_{\rm TM}$	& TM relation: Normalization 						& $3.02 \times 10^{14} \MSUN \HI$	& $\mathcal \delta_D (3.02 \times 10^{14} \MSUN \HI) $ \\
$\sigma_{\rm TM}$ 	& TM relation: Logarithmic Scatter 					& 0.119 						&$\mathcal \delta_D (0.119) $  \\ 

 & & & \\
\hline
\end{tabular}
\end{center}
\end{table*}

\section{Cosmological models}
\label{SEC:SCENARIOS}
Our reference cosmological model ($\Lambda$CDM) is a spatially flat, Gaussian, vanilla, Friedmann-Lemaitre-Robertson-Walker universe, described by 5 parameters: 3 for the background -- $\Omega_{\rm m}$, $\Omega_{\rm b}$, and $h$ --  and 2 for the scalar perturbations -- $n_{\rm s}$ and $\sigma_8$. Here, 
DE coincides with the cosmological constant $\Lambda$ and is modelled as a fluid whose equation of state (EOS), $ p = w \rho c^2$, is time independent with $w=-1$.
Throughout the paper, the flatness assumption translates into a condition for the DM and DE densities, $\OL = 1 - \OM$, and we consider Gaussian perturbations, i.e. $\FNL=0$.
We pick a fiducial model (Table~\ref{TAB:PARAM}) by adopting the best-fitting values for the cosmological parameters from \cite{Komatsu:2009} based on the combination of 5-yr data from the Wilkinson Microwave Anisotropy Probe ({\it WMAP}), baryonic acoustic oscillations (BAO), and supernovae Ia (SN). 
As it will become clear later on, this choice, which is congruous with the 9-yr results by \cite{Hinshaw:2013}, is motivated by the requirement of achieving internal consistency in our calculations.

We investigate the nature of DE by either allowing the EOS parameter $w$ to deviate from $-1$ in a time-independent fashion ($w_0$CDM), or by modelling
DE as a dynamical fluid ($w$CDM).
In the latter case, we adopt
the phenomenological parameterization for $w$ used by the Dark Energy Task Force \citep[DETF,][]{Albrecht:2006}, $w(a) = w_0 + (1-a)\,w_a$, where $a$ is the time-dependent scale factor of the Universe, $w_0$ is the present-day value of the EOS parameter, and  $w_a$ is its rate of change \citep{Chevallier:2001, Linder:2003}. We assume that the sound speed  coincides with the speed of light (this class of models includes all the canonically coupled scalar fields) and therefore neglect the effect of DE on the cold DM power spectrum.
In Table \ref{TAB:COSMO_SCEN}, we summarize and characterize the cosmological scenarios investigated in this work, together with the list of the cosmological parameters adopted for the marginalization.

\begin{table}
\begin{center}
\caption{\label{TAB:COSMO_SCEN} Cosmological models studied in this paper and cosmological parameters adopted for the marginalization in addition to the four X-ray luminosity--mass scaling-relation parameters, that are always allowed to vary in the statistical analysis, yet with various prior strategies (see Table \ref{TAB:PARAM} and Section \ref{SEC:SELFCAL}). Throughout the paper, we assume a flat cosmology, with $\OL = 1 - \OM$, and external priors on $h$ and $\OB$ described in Section \ref{SEC:SCENARIOS}.}    
\begin{tabular}{l l l }
\hline
&&\\
Model 					& Symbol					& Parameter Set (+ LM Sector)											\\
&&\\
\hline
&&\\
Vanilla Model				& $\Lambda$CDM 			& \{${\SIGMA8, \OM, n_{\rm s}, h, \OB}$\}							 \\
Constant DE 				& $w_0$CDM 				& \{${\SIGMA8, \OM, n_{\rm s}, h, \OB, w_0}$\}						\\
Evolving Models 			& $w$CDM   				& \{${\SIGMA8, \OM, n_{\rm s}, h, \OB, w_0, w_a}$\}					\\
&&\\
\hline
\end{tabular}
\end{center}
\end{table}


\section{Setup, Definitions, and Methods}
\label{SEC:SETUP}

We study the impact of DE on the mean number density and spatial distribution of the galaxy clusters which will be detected by $\eRO$. Such influence derives from the fact that the properties of DE alter the expansion history of the Universe and, as a consequence, the growth rate of structure and the distance-redshift relations.
Consistently with the adopted parameterization for the DE EOS, the evolution of the Hubble factor can be written as
\BE
E(z) = \frac{H(z)}{H_0} = \sqrt{ \OM(1+z)^3 + \OL f_{\rm DE}(z)}
\EE
where $H_0 = 100\, h $ km ~s$^{-1}$ Mpc$^{-1}$ and 
\begin{eqnarray}
f_{\rm DE}(z) &=& \exp \left[  -3 \int_1^{a(z)} \frac{1+w(a')}{a'}\right] {\rm d}a' \nonumber\\
&=& a^{ - 3(1+w_0+w_a)}~ \exp \left[ -3 w_a(1-a)\right].
\end{eqnarray}
%
The linear growth factor for matter-density perturbations on sub-horizon scales, $D_+$,
can be determined by solving \citep[e.g.][]{Percival:2005,Linder:2003}:
\BE
D_+'' + \left( \frac{3}{a} + \frac{{\rm d} \log H}{ {\rm d}a} \right) D_+' - \frac{3 \OM}{2 a^5 E^2(a)}D_+ = 0\;,
\label{eq:D+}
\EE
where, with a little abuse of notation, we denote by $E(a)$ the
composite function $E[z(a)]$ with $z(a)=a^{-1}-1$.
For particular DE models, it is possible to write analytical solutions (or accurate approximations) for $D_+$ 
\citep[e.g.][for a cosmological constant and a flat universe with constant $w$, respectively]{Heath:1977, Percival:2005}. 
However, here we solve equation (\ref{eq:D+}) numerically to properly accommodate for DE models with an evolving equation of state.\\

We calculate galaxy-cluster number counts and power spectra by following the method introduced in \cite{Pillepich:2012} and briefly summarized below.
$\eRO$ galaxy clusters are selected in terms of the raw photon counts\footnote{This improves upon common but unrealistic approaches in the literature that either select clusters based on their un-observable virial mass or 
assume a generic mass-observable relation.} ($\eta$) that will be collected at the detectors. 
To do so, we follow a series of steps. First, we compute the mass function and the linear bias coefficient of dark-matter haloes following \cite{Tinker:2008} and \cite{Tinker:2010a}. Subsequently, we associate each halo with an X-ray flux by adopting observationally motivated scaling relations, namely the luminosity--mass and temperature--mass relations from \cite{Vikhlinin:2009a} (see Table \ref{TAB:PARAM}). We allow the four parameters of the luminosity--mass relation to vary throughout the statistical analysis, unless otherwise stated. We therefore take into account: (i) the spectral energy distribution of the ICM emission \citep[{\it apec} in {\sc xspec:}][]{Smith:2001, Arnaud:1996} by assuming an average intracluster metallicity of 0.3 $Z_{\odot}$ \citep{Anders:1989};
(ii) the photoelectric absorption suffered by the photons along the line sight by assuming an average hydrogen column density of $3 \times 10^{20}$ atoms cm$^{ -2}$ \citep{Kalberla:2005}; (iii) the expected telescope response (by Frank Haberl: erosita\_iv\_7telfov\_ff.rsp). Eventually, we are able to evaluate one- and two-point clustering statistics as a function of the raw photon counts.

It is important to mention that the linear matter transfer function has been computed 
using the {\sc camb}\footnote{http://camb.info/, with default choices but for the adjustments to our fiducial cosmological parameters and selecting `High precision: Yes', `Matter/Power: Calculated Values', `k per logint' = 10.} code \citep{Lewis:2000} and that, throughout the text, cluster masses are defined in terms of the spherical-overdensity criterion with respect to the critical density of the Universe, i.e. $\M500c = (4 \pi /3)\, \Delta_{\rm crit}\, \rho_{\rm crit}(z)\, R_{\rm 500c}^3$ with $\Delta_{\rm crit} \equiv 500$ and $R_{\rm 500c}$ being the radius encompassing a sphere with average matter density equal to 500 times the critical density of the Universe $\rho_{\rm crit}(z)$.
%

\subsection{Survey strategies and parameters}
$\eRO$ will perform an X-ray all-sky survey in about 4 years, after reaching a L2 orbit. The entire celestial sphere will be scanned in 8 successive passages, with overlap of all great circles at the ecliptic poles and a relatively uniform coverage of the entire sky at progressively deeper exposures. In the following, $\eRA$ stands for $\eRO$ all-Sky Survey and the numbers 1-8 denote increasing exposure from the first to the eighth passage. 

In Table \ref{TAB:SURVEY_PARAM}, the survey parameters relevant for the analysis are summarized in terms of energy band, sky fraction, exposure time, photon count detection threshold, and an additional mass cut: this is a redshift-dependent photon-count threshold aimed at removing too low-mass objects that would pass the photon threshold at low redshifts \citep[see][for details]{Pillepich:2012}. Although cluster detectability depends on a host of factors, including e.g. the extent of the source, the adopted 50-photons limit as detection threshold has been demonstrated to be a reliable zeroth-order choice (see \textcolor{blue}{Clerc et al., 2018}). With the adopted fiducial cosmological parameters and the values reported in Table \ref{TAB:SURVEY_PARAM}, we expect $\eRO$ to detect $\sim 8.89 \times 10^4$ clusters of galaxies ($\sim 9.32 \times 10^4$, including Poisson errors in the photon counts) with a median redshift of $\sim$ 0.35, when objects below $5\times 10^{13}h^{-1} \MSUN$ ($\sim 7 \times 10^{13} \MSUN$) are removed. Such numbers correspond to a maximum redshift of $z\sim2.5$, while only about $10^3$ clusters are expected to reside beyond $z\gtrsim 1$.
The all-sky data will be split into two equal, non-overlapping parts between the German and Russian Consortia, respectively: however, in this paper, we will consider a sky fraction of $\sim$ 66 per cent as in \cite{Pillepich:2012} but provide results for half the sky in Section \ref{SEC:ERASSn} for completeness. 

After the completion of the all-sky survey, pointed observations will be undertaken for other 3.5 years. Spectroscopic redshifts will be readily available thanks to the detection of iron lines in the cluster spectra: yet, this will be feasible only for a small subsample of $\eRO$ clusters \citep{Yu:2011, Borm:2014}. Optical and near-infrared follow-up campaigns are being planned to complement the $\eRO$ data with spectroscopic redshifts, at least up to $z \sim 0.7$. These include a survey with the new optical multi-object spectrograph 4MOST\footnote{4-m Multi-Object Spectroscopic Telescope for ESO} at the ESO VISTA telescope \citep{deJong:2014, Depagne:2014},
as well as SPIDERS\footnote{SPectroscopic IDentification of ERosita Sources} and Black Hole Mapper, 
conducted by SDSS IV/eBOSS and SDSS V, respectively.
Conservatively speaking, spectroscopic 
redshifts for clusters can be measured with an accuracy of
$0.0025(1+z)$, with SPIDERS aiming at $0.001(1+z)$ out to $z\sim 0.6$ \citep{Clerc:2016}. In parallel, a host of multi-band photometric data will be used to identify cluster and AGN counterparts, and to estimate photometric redshifts. Overlap of the $\eRO$ sky with completed or ongoing programs like SDSS, PanSTARRS\footnote{Panoramic Survey Telescope and Rapid Response Systems}, and DES\footnote{Dark Energy Survey} will provide cluster redshifts with accuracies ranging from $0.018(1+z)$ at $z \lesssim 0.6$ to $0.025(1+z)$ up to at $z \sim 1.1$ \citep{Merloni:2012} or even better \citep[e.g. $\lesssim 0.02(1+z)$ all the way up to $z \lesssim 1.3$,][]{Drlica-Wagner:2018}. 
\begin{table}
\begin{center}
\caption{\label{TAB:SURVEY_PARAM} $\eRO$ All-Sky Survey ($\eRA$) parameters, for 65.8 per cent sky coverage excising $\pm 20~ {\rm deg}$ around the Galactic plane. $\eRO$ will scan the sky eight times in four years, with an average integrated exposure time of 1.6 ks.}
\begin{tabular}{l c }
\hline
&\\
Choices/Description 					& {\it eRASS:n}  		\\
&\\
\hline
&\\
X-ray Energy Band [keV]                  & 0.5-2.0                     \\							                          
Detection limit (min raw photon count): $\eta_{\rm min}$  		& 50 				\\
Minimum cluster mass [$h^{-1}\MSUN$] 	& $1 ~{\rm or}~ 5\times10^{13} $\\
Sky coverage: $f_{\rm sky}$  [${\rm deg}^2$]				&  27,145					\\
Exposure Time: $T_{\rm exp}$ [s]						& $n \times 200$\\

&\\
\hline
\end{tabular}
\end{center}
\end{table}

\subsection{Probes}
We consider two experimental probes: the number counts of galaxy clusters as a function of redshift and their angular clustering that we quantify in terms of the power spectrum.
Unless otherwise stated, throughout the paper we will quote
cosmological constraints obtained combining the two probes.
We refer the reader to \cite{Pillepich:2012} for a detailed description of our method while here we only briefly summarize its most important features and how we model the different observables. 

\subsubsection{Number counts}
The total number of clusters detected above a certain photon-count threshold and within a given redshift bin can be calculated by integrating the redshift distribution of the clusters:

\BE
\frac{{\rm d}N_{> \eta_{\rm min}}}{{\rm d}z} (z) = 4 \pi f_{\rm sky} \left[\frac{c}{H(z)} D_{\rm A}^2(z)\right] 
\int^{\infty}_{\eta_{\rm min}} \frac{{\rm d}n}{{\rm d}\eta}(\eta, z) ~ {\rm d}\eta,
\label{EQ:DNOVERDZ}
\EE
where 
$c$ is the speed of light, $D_{\rm A}$ is the comoving angular diameter distance, and $ \frac{dn}{d\eta}(\eta, z)$ denotes the DM halo mass function properly mapped into a raw-count function for the $\eRO$ clusters.

\subsubsection{Angular clustering}
After splitting the clusters in bins of $\eRO$ photon counts and redshift,
we adopt the Limber approximation to compute their angular (auto and cross) power spectra,
\BE
C_\ell(i,j) = 4\pi \int_0^\infty {\rm d}z~ \frac{{\rm d}V}{{\rm d}z} ~P_{\rm lin}
\left(\frac{\ell+1/2}{D_{\rm A}},z\right)  ~W_i(z) \,W_j(z)\;.
\label{EQ:ANGCLUSTERINGL}
\EE
Here, $V$ is the comoving volume and
$P_{\rm lin}(k,z)$ denotes the linear matter power spectrum evaluated at wavenumber $k$ and redshift $z$.
The weight function
for the $i$-th bin is
\BE
W_i(z) = \frac{1}{N_i} ~\frac{{\rm d} N_i}{{\rm d}V}(z) \,b_i(z)\;,
\EE
where $N_i$ gives the number counts of the clusters and 
$b_i$ is their linear bias parameter. 
We consider spherical harmonics with
$5<\ell\leq\ell_{\rm max}$. The minimum
value has been set considering
the wide-area extension of the survey.  On the other hand, we fix
$\ell_{\rm max}$ based on the angular separation which corresponds to a wavenumber of $k_{\rm max} = 0.1 ~h \,$Mpc$^{-1}$, in order to alleviate potential issues related to cluster exclusion effects as well as non-linearities in the dynamics of the density perturbations and in the cluster bias (see  Section 6 in \citealt{Pillepich:2012}). Note that, although we partition the clusters into a number of redshift bins, we do not take into account the cross-spectra between the angular positions of clusters in different redshift slices.
In this work, $C_\ell(i,j)$ thus indicates the angular spectrum between photon-count bins within the same redshift interval.

\begin{figure*}
\BC
\includegraphics[width=17cm]{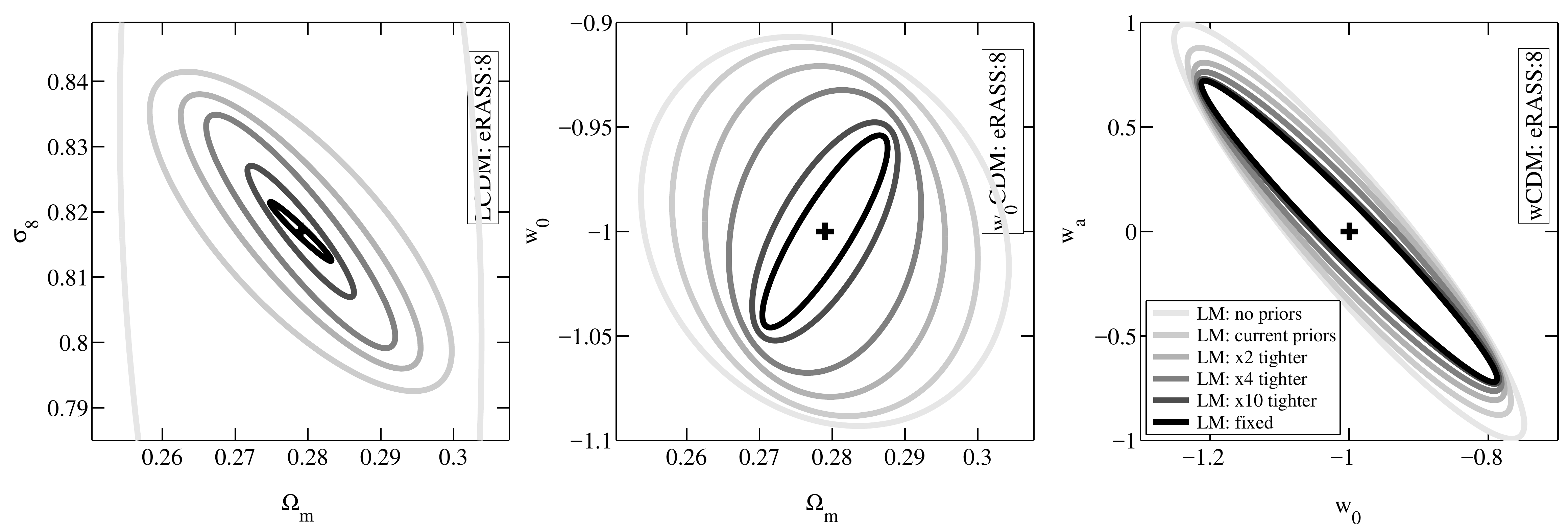}
\caption{\label{FIG_PRIORS} The impact on the cosmological forecast for $\eRO$
of adopting external constraints on the X-ray luminosity--mass scaling relation.
The label `no priors' corresponds to the self-calibration technique; `current priors' refers to the relative errors measured by \citet{Vikhlinin:2009a}, while the tags `$\times N$ tighter' indicate the cases in which those can be improved by a factor of $N$. The limiting situation in which the parameters are perfectly known is indicated with `fixed'.
The three panels refer to different cosmological scenarios as indicated by the labels. 
Results assume four years of observations and are obtained from the combination of cluster counts (with photometric redshifts) and clustering.
}
\EC
\end{figure*}

\subsection{Figure of merit}
\label{SEC:FISHER}
The relative performance of different experiments in constraining a subset of model parameters can be conveniently quantified by introducing a figure of merit (FoM) which is inversely
proportional to the volume of their 68.3 or 95.4 per cent joint credible region.  
The higher the FoM, the more suitable an experiment is to constrain the selected parameter set.
We define a FoM for the DE parameters $w_0$ and $w_a$ in terms of their (marginalized) covariance matrix as
\BE
{\mathrm{FoM^{\rm DE}}} = \left[ \det{\mathrm{Cov}}( w_0, w_a)\right]^{-1/2}\;.
\label{EQ:FOM_DETF}
\EE
In order to
compute $\mathrm{Cov}( w_0, w_a)$
we first invert the full
Fisher-information matrix and then extract the $2\times2$ submatrix corresponding to the DE parameters.
Since the Fisher-matrix formalism assumes a Gaussian posterior distribution, our FoM is a factor $6.17 \pi$ larger than that introduced by the DETF \citep{Albrecht:2006} but exactly matches 
the entries $[\sigma(w_a)\times \sigma(w_{\rm p})]^{-1}$ in the tables of their Section IX (here $w_{\rm p}$ denotes the value of $w$ at a pivot redshift chosen so that errors in $w_a$ and $w_{\rm p}$ are uncorrelated)\footnote{The area enclosed by the joint 68.3 and 95.4 per cent credible intervals of $(w_0,w_a)$  is $\alpha\,\pi\,\left[ \det{\mathrm{Cov}}( w_0, w_a)\right]^{1/2}$ with $\alpha=2.30$ and 6.17, respectively. 
Although the DETF formally defines the figure of merit for DE as the reciprocal of the area of the 95.4 per cent credible interval, in the tables of their report they quote results obtained setting $\alpha \pi=1$ as in our equation~(\ref{EQ:FOM_DETF}).}.

Throughout the paper,
$\sigma(p)$ indicates the forecast uncertainties on the parameter $p$
and gives the rms error obtained after marginalizing over all the other model parameters.
Similarly, we visualize
constraints for variable pairs by plotting the boundary of their joint
68.3 per cent credible region (after marginalizing over the remaning parameters).
We present results and quote FoM values for both $\eRO$ alone as well as $\eRO$ in combination with other data. In fact, all values in the tables of Section IX of \citet{Albrecht:2006} give 
 forecasts for future probes {\it combined with {\it Planck}}. Hence, only analog combinations of $\eRO$ data with {\it Planck} priors can be compared with them.


\section{Binning Strategy}
\label{SEC:BINNING}
We split the $\eRO$ cluster sample into bins both in photon counts and redshifts. 
We explore various options to choose the optimal binning strategy that maximizes the return of the data set.

For the photon counts, we consider either equal-sized logarithmic bins in the range $50 \lesssim \eta \lesssim 5\times 10^4$ or variable-sized bins approximately containing the same number of clusters. 
For the redshift, we take bins of size
$\Delta z \,(1+z_{\rm bin})$, where $\Delta z$ is a parameter we vary
and $z_{\rm bin}$ is the central value in a bin.
We consider $\Delta z = 0.2, 0.1, 0.05, 0.02, \text{and } 0.01$, corresponding to 5, 10, 21, 54, and 108 slices in the range $0.01<z\lesssim 2$.
In what follows, we refer to the cases with $\Delta z = 0.05$ and $\Delta z = 0.01$ as
`photometric' and `spectroscopic' accuracies, respectively. This conservative designation refers
to the fact that such bin widths correspond to the typical $3--4\sigma$ redshift errors expected for the $\eRO$ clusters (see Section \ref{SEC:SETUP}).

We find that, in order to break the strong degeneracies among the model parameters,
it is necessary to finely sample the data both in redshift and in photon counts, $\eta$.
In general, for a fixed total number of bins, binning in redshift is more efficient than binning in the photon counts. However, the latter comes at no cost and should always be applied. 

The results we quote in the paper for the cluster counts are well converged by considering 20 bins in $\eta$ (logarithmic or not: the forecast does not depend on this) and assuming either photometric or spectroscopic redshift accuracies. 
On the other hand, for the clustering measurement, we adopt a different binning strategy.
In fact, while the angular power-spectrum grows by considering thinner and thinner redshift slices, the number of clusters per bin rapidly drops and the systematic shot-noise correction overcomes the actual clustering signal. If shot noise was exactly Poissonian, it would not be a problem to subtract its contribution out before fitting the power spectra with a model. However, in reality, it is not obvious how to accurately model discreteness and exclusion effects. Therefore, we prefer not to work in a regime where shot noise dominates, which happens
already whenever we consider about 30 cells in total.
%
For the clustering study, we therefore first slice the data into four redshift bins spanning the range $0.1<z\lesssim 2$  and further partition them into five $\eta$ bins chosen so that $\sim 4,000$ clusters are assigned to each of them. Note that this binning strategy
is not particularly demanding in terms of redshift-measurement errors as it roughly corresponds to setting $\Delta z\sim 0.1$.
We have checked that our results are well converged once clustering and cluster counts are combined together.


\section{Beyond self-calibration: \\external priors on the X-ray sector}
\label{SEC:SELFCAL}
In principle, we could use the $\eRO$ data to simultaneously constrain cosmology, selection effects, and the cluster mass-observable relation. As in \citet{Pillepich:2012}, this could be achieved by adopting very broad non-informative priors on the slope, normalization, time-evolution and scatter of the X-ray luminosity--mass relation ($\ALPHALM, \BETALM, \GAMMALM$ and $\SIGMALM$). However, such self-calibration scheme would provide conservative results as, 
in effect, some information on the X-ray scaling relation is already available. Even more importantly, follow-up observations are already being planned to provide subsets of $\eRO$ clusters with more stringent mass estimates. 

In this Section, we quantify the impact of adopting informative priors on the 
parameters of the $L_X-M_{500}$ relation.
We use the functional form given by \cite{Vikhlinin:2009a}, who also measured 1$\sigma$ uncertainties on the best-fit parameters from two sets of {\it Chandra} clusters with median redshifts of about 0.05 and 0.5. 
The corresponding relative errors on $\ALPHALM, \BETALM, \GAMMALM$ and $\SIGMALM$ are 8.6, 23, 0.1, 9.8 per cent, respectively, with negligible covariances for the purposes at hand (if the scaling relations are written as in equation 18 of \citet{Pillepich:2012}; Vikhlinin, private communication).
In Fig. \ref{FIG_PRIORS}, we show how the constraints on the cosmological parameters set by $\eRO$ change when the current uncertainties on the scaling-relation parameters
will be improved by a factor of $N\geq 1$ thanks to the synergy of X-ray follow up with Chandra, XMM-Newton, Suzaku, possibly NuSTAR\footnote{The Nuclear Spectroscopic Telescope Array \citep{Harrison:2013}}, and $\eRO$ itself. 

It is apparent that considering external information on the scaling relations is key to get tighter cosmological constraints. This remains true also when data from other probes are considered (e.g. combining $\eRO$ with {\it Planck}, not shown in the figure). In particular, we find that the improvements due to the refinement of the LM sector are much more pronounced for $\SIGMA8$ and $\OM$ than for the DE parameters (since the correlations between the LM and the DE sectors are weaker). For instance, considering an evolving DE model (Fig.~\ref{FIG_PRIORS}, right panel) and comparing the results obtained assuming $N=4$ with those of the self-calibration, we find that the uncertainties on $\SIGMA8$ and $\OM$ shrink by a factor of 4 and 2, respectively, while those on $w_0$ and $w_a$, in comparison, {\it only} reduce by 20--30 per cent.
Further discussion about the dependence of our forecast on the adopted priors for the LM relation will be presented in the next Section.


\section{Results}
\label{SEC:RESULTS}
We present here the results of our forecast. We refer to the
full $\eRO$ sample, obtained after four years of observations ({\it eRASS}:8). Unless otherwise stated, we quote marginalized 68.3 per cent credible intervals derived from the combination of number counts and angular clustering. 

As summarized in Table \ref{TAB:ERO_ANALYSIS}, we focus on two scenarios. We call `pessimistic' the case where (i) photometric redshifts are available for all the clusters, (ii) the size of the
adopted priors for the LM sector coincides with the measurement errors in
 \cite{Vikhlinin:2009a}, and (iii) the functional form and accuracy of the LM scaling-relation can be trusted down to cluster masses of $\M500c = 5 \times 10^{13} h^{-1} \MSUN$. 
On the other hand, we call `optimistic'\footnote{Note that our strategy is somewhat more conservative than that adopted by the DETF for cluster data: there, an improvement by a factor of 7 is assumed in the mass-observable relation parameterization between their pessimistic and optimistic scenarios, with an optimistic relative uncertainty on the mean and variance per redshift bin as small as 1.6 per cent. However, in both StageIII and StageIV experiments, the DETF assumed surveys with a much smaller number of detected clusters than the actual $\eRO$ cluster count.}  the case with (i) spectroscopic redshifts, (ii)  priors on the LM relation which are improved by a factor of $N=4$ compared to \citet{Vikhlinin:2009a}, and (iii) the inclusion in the analysis of all 125,300 clusters 
with $\eta\geq 50$ and $\M500c \gtrsim 1 \times 10^{13} h^{-1} \MSUN$.
In this second option, we extrapolate the functional form (and fiducial values) of the LM relation to galaxy groups.
In fact, X-ray studies which extend scaling relations towards masses of $\sim10^{13} h^{-1} \MSUN$ are available \citep{Sun:2009, Eckmiller:2011OK, Lovisari:2015, Bharadwaj:2015}, and have shown both an increase in the scatter \citep{Eckmiller:2011OK} as well as a steepening of the LM relation \citep{Lovisari:2015}. However, our goal here is to emphasize the impact of low-mass objects and, for this reason, we opt for a simplified approach.

Our baseline cluster analysis includes, in addition to the {\it eRASS}:8 dataset and mass proxies from X-ray follow-up data, Gaussian priors on the Hubble parameter 
\citep[with $\Delta h = \pm 0.022$,][]{Riess:2011} and on the cosmic baryon density \citep[with $\Delta (\OB h^2) = \pm0.0045$,][]{Cooke:2014a}.
%

\begin{table*}
\begin{center}
\caption{\label{TAB:ERO_ANALYSIS}Scenarios for the analysis of the $\eRO$ data. `Self calibration' denotes the configuration adopted for the main results of {\color{blue}{Pillepich \etal 2012}}, i.e. when {\it no priors} are applied to the X-ray observables -- mass relation parameters. `Photo-z' and `spectro-z' redshift accuracies are implemented here in terms of the width of the redshift slices the data can be binned into, i.e. of width $\Delta z\,(1+z)$ with $\Delta z = 0.05$ and $\Delta z = 0.01$, respectively. This distinction is to be intended only for the counts experiment, as for the clustering the slicing in redshift-space is limited by shot-noise to a handful of bins in redshift space. The pessimistic and optimistic scenarios which we refer to throughout the paper also differ according to the strength of the priors which can be placed on the observable-mass relation parameters and according to the minimum cluster mass which can be reliably modeled and therefore included in the analysis.}
\begin{tabular}{l c c c c}
\hline
&& & &\\
Scenario 			& Priors on LM parameters 			& Redshift Bins Accuracy  			& Minimum Cluster Mass   & \# Clusters\\
&& & $M_{\rm 500c}$ [$\MSUN/h$]&\\
&& &&\\
\hline
&& &&\\
Self calibration 	& $\mathcal{U(-\infty,+\infty)}$ 			& photo-$z$: $\Delta z = 0.05$ 			& $5\times 10^{13}$			& 88,900\\
Pessimistic 		& current knowledge		 			& photo-$z$: $\Delta z = 0.05$			& $5\times 10^{13}$			& 88,900\\
Optimistic 			& x 4 better than current knowledge 		& spectro-$z$: $\Delta z = 0.01$		& $1\times 10^{13}$			& 125,300\\
&& &&\\
\hline
\end{tabular}
\end{center}
\end{table*}

\begin{figure*}
\BC
\includegraphics[width=15.8cm]{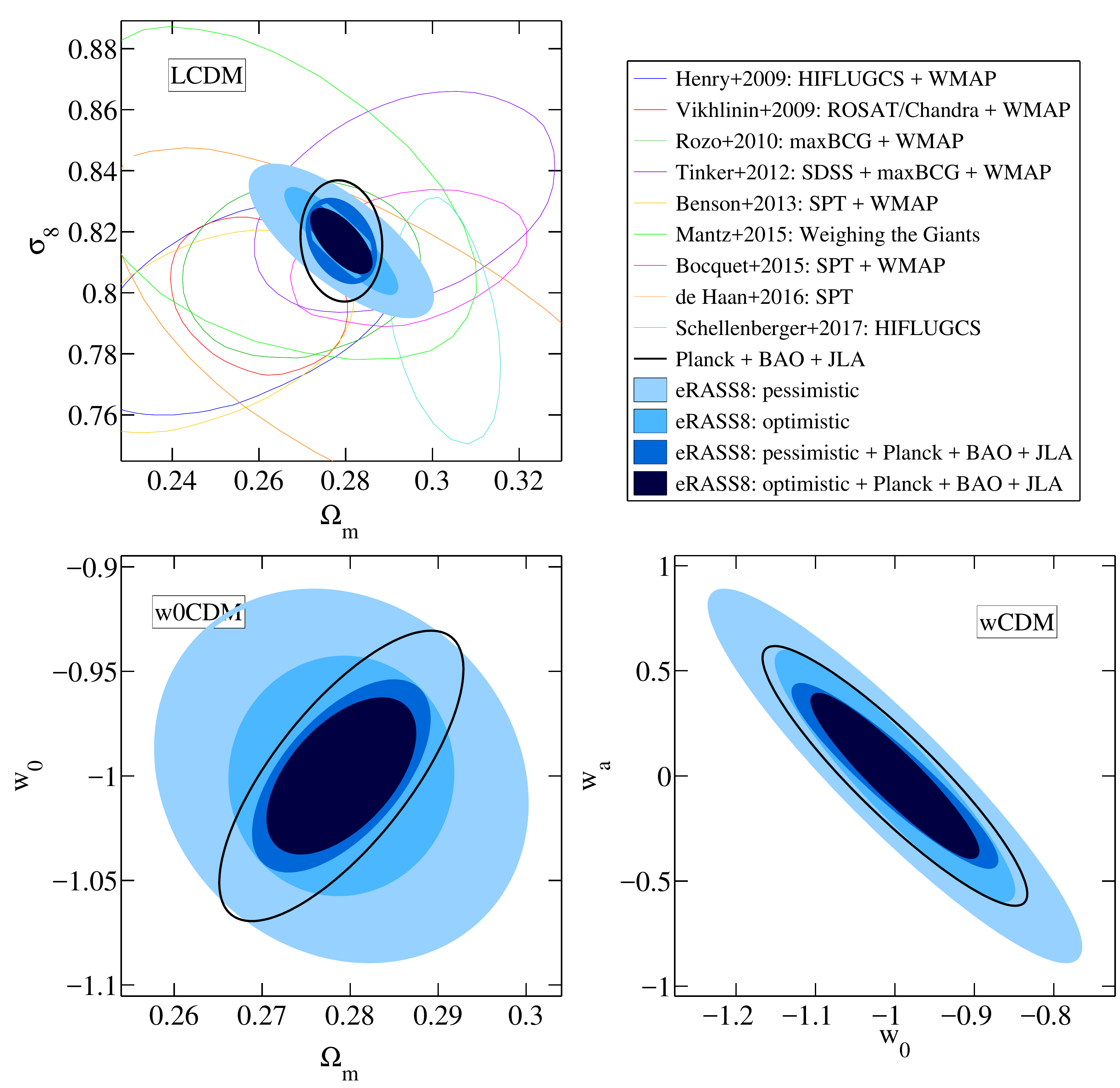}
\caption{\label{FIG:FM_MAINRESULTS} $\eRO$ forecasts after 4 years of all-sky survey. Joint 68.3 per cent credible regions are shown is shades of blue for a selection of parameter pairs, from the combination of cluster abundances and clustering, obtained by marginalizing over all the other model parameters (Table \ref{TAB:COSMO_SCEN}), in a $\Lambda$CDM cosmological model (top), in a cosmological model with constant (bottom left) and evolving (bottom right) DE equation-of-state parameters, respectively. Thin color curves denote analog constraints from other cluster samples, in combination or not with additional data. Black thin contours display the results from the combination of {\it Planck}, BAO and supernova data -- see text for details.}
\EC
\end{figure*}

\begin{table*}
\begin{center}
\caption{\label{TAB:RESULTS} $\eRO$ forecasts after 4 years of all-sky survey: one-dimensional, 1-$\sigma$ errors and Figure of Merit corresponding to the contours of Fig.~\ref{FIG:FM_MAINRESULTS} and obtained by marginalizing over the parameters listed in Table \ref{TAB:COSMO_SCEN}. The dark-energy figure of merit is defined in Eq.~\ref{EQ:FOM_DETF}.}
\begin{tabular}{l l l c c c c c}
\hline
&&&&&&\\
Data	& Scenario 	&  Model 	& $\Delta\SIGMA8$ & $\Delta\OM$ & $\Delta w_0$ & $\Delta w_a$ &FoM$^{\rm DE}$ \\
&&&&&\\
\hline

\hline
&&&&&\\

{\it eRASS}:8 + $H_0$ + BBN							& Pessimistic   	& $\Lambda$CDM 			& 0.016 		& 0.014 &- & - & -  \\ 
{\it eRASS}:8 + $H_0$ + BBN							& Optimistic    	& $\Lambda$CDM			& 0.011 		& 0.008 &- & - & -  \\ 
{\it eRASS}:8 + $H_0$ + BBN+ {\it Planck} + BAO + JLA		& Pessimistic   	& $\Lambda$CDM 			& 0.009 		& 0.005 &- & - & -  \\ 
{\it eRASS}:8 + $H_0$ + BBN+ {\it Planck} + BAO + JLA		& Optimistic    	& $\Lambda$CDM 			& 0.007 		& 0.004 &- & - & -  \\ 
 
&&&&&\\
{\it eRASS}:8 + $H_0$ + BBN							& Pessimistic   	& $w_0$CDM 				& 0.017 		& 0.014	 		& 0.059&  - & - \\ 
{\it eRASS}:8 + $H_0$ + BBN							& Optimistic    	& $w_0$CDM 				& 0.011 		& 0.008 		& 0.037&  - & - \\ 
{\it eRASS}:8 + $H_0$ + BBN + {\it Planck} + BAO + JLA 	& Pessimistic   	& $w_0$CDM 				& 0.010 		& 0.007 		& 0.030&  - & - \\ 
{\it eRASS}:8 + $H_0$ + BBN + {\it Planck} + BAO + JLA 	& Optimistic    	& $w_0$CDM 				& 0.007 		& 0.005 		& 0.024&  - & - \\ 

&&&&&\\
{\it eRASS}:8 + $H_0$ + BBN							& Pessimistic   	& $w$CDM 				& 0.029 		& 0.022 		& 0.154			& 0.58 	& 29 \\ 
{\it eRASS}:8 + $H_0$ + BBN							& Optimistic    	& $w$CDM 				& 0.015 		& 0.010 		& 0.098 		& 0.39 	& 69 \\ 
{\it eRASS}:8 + $H_0$ + BBN + {\it Planck} + BAO + JLA 	& Pessimistic   	& $w$CDM 				& 0.011 		& 0.007 		& 0.084 		& 0.29 	& 116 \\ 
{\it eRASS}:8 + $H_0$ + BBN + {\it Planck} + BAO + JLA 	& Optimistic    	& $w$CDM 				& 0.008 		& 0.006 		& 0.068 		& 0.25	& 162 \\ 

&&&&&&\\
\hline
\end{tabular}
\end{center}
\end{table*}

\subsection{Constraints from {\it eRASS:8}}
\label{SEC:RESULTS_eRASS8}
The constraints that $\eRO$ will set on the cosmological parameters with its 4-yr all-sky survey are reported for a subset of relevant parameters in Table \ref{TAB:RESULTS} and Fig.~\ref{FIG:FM_MAINRESULTS} considering
three different cosmological models: $\Lambda$CDM, $w_0$CDM, and $w$CDM (see Table \ref{TAB:COSMO_SCEN})\footnote{Constraints from $\eRO$ for all parameters included in the marginalization are available upon request.}. 

In the upper panel of Fig.~\ref{FIG:FM_MAINRESULTS}, the ability of $\eRO$ to constrain
$\SIGMA8$ and $\OM$ within a $\Lambda$CDM model is compared to the results of a variety of complementary cluster measurements from the literature (solid thin colored contours), via optical  \citep{Rozo:2010, Tinker:2012}, X-ray \citep{Henry:2009, Vikhlinin:2009b, Mantz:2015,Schellenberger:2017}, and Sunyaev-Zel'dovich observations \citep{Benson:2013, Bocquet:2015a, deHaan:2016}, all combined but for three cases to CMB results available at that time\footnote{The combination with CMB data is manifest in the inclination of the elliptical contours: cluster data alone provide anti-correlated estimates for $\SIGMA8$ and $\OM$ while {\it WMAP} data give correlated measurements. The cluster samples quoted from the literature are made by a few tens to a few hundreds of clusters, and indeed their CMB-combined cosmological constraints are often strongly dominated and determined by CMB data. Note that \citet{Schellenberger:2017} combined the constraints from the cluster mass function with constraints from the cluster gas mass fraction, which shifted the best fit of $\OM$ to a larger value, within the uncertainties of the cluster-count only results (see their Fig. 1).}. Owing to its $10^5$ clusters, $\eRO$ {\it alone} will largely improve upon any currently available cluster dataset, with constraints on $\SIGMA8$ and $\OM$ of the order of 1 per cent and 3 per cent, respectively, even when a constant DE equation-of-state parameter is included. 

In the case of spatially flat, constant-$w$ models, constraints on $\SIGMA8$ and $\OM$ from $\eRO$ alone are identical to the flat $\Lambda$CDM ones, with $\Delta w_0$ as good as $\pm 0.037$. For models with an evolving equation of state, $\eRO$ alone will be able to provide marginalized, one-dimensional, 1$\sigma$ errors as good as $\Delta \SIGMA8 = \pm~0.015$, $\Delta \OM = \pm~0.010$, $\Delta w_0 = \pm~0.098$, and $\Delta w_a = \pm~0.39$ (optimistic scenario). These results correspond to a DE FoM of about 69 ($\eRO$ alone).

\begin{figure*}
\BC
\includegraphics[width=15.6cm]{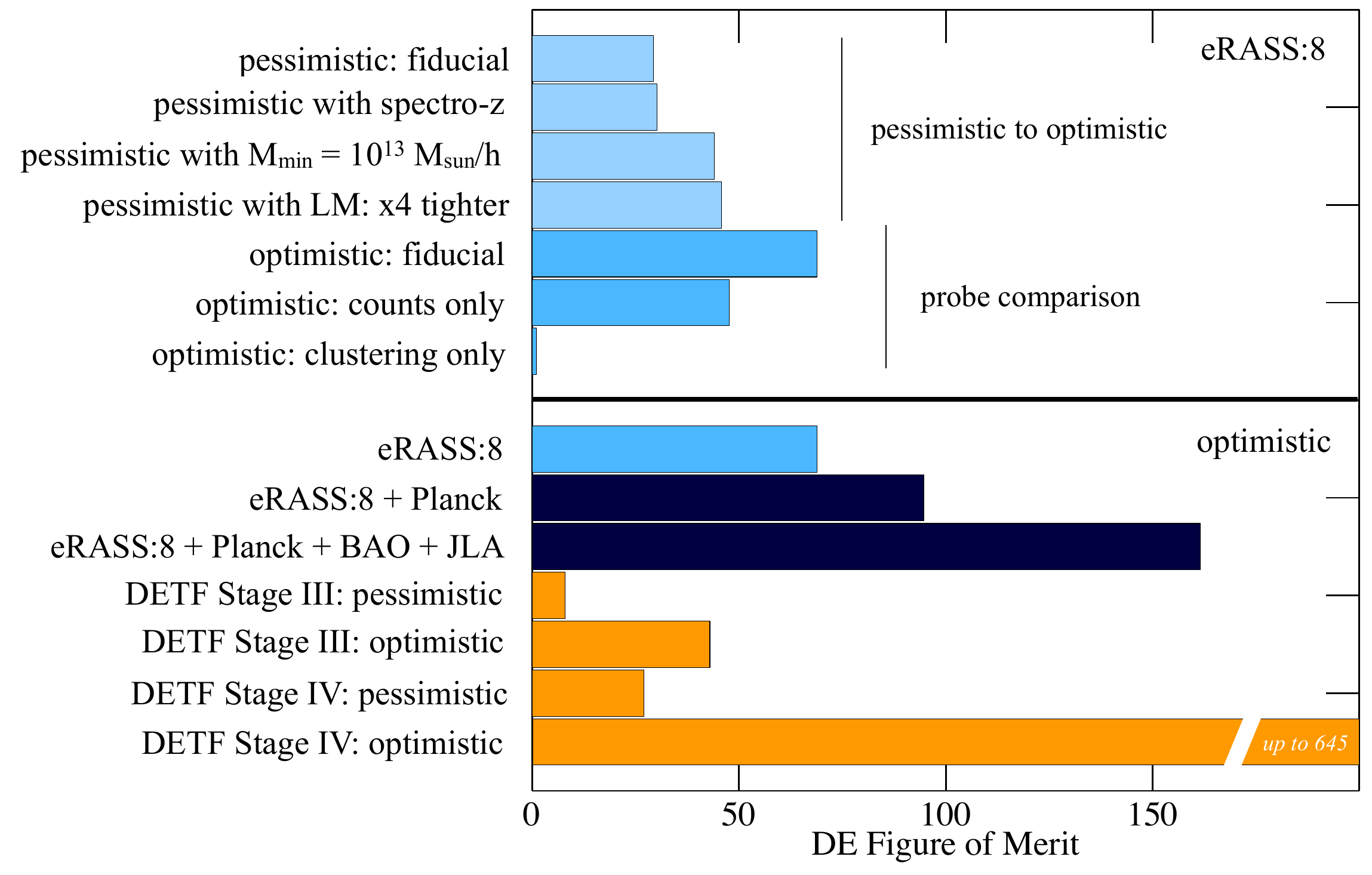}
\caption{\label{FIG:FM_COMP} The DE FoM from $\eRO$ for different survey assumptions and configurations (top) and in comparison to the DETF expectations (bottom). In the top panel, we highlight the impact of individual choices in tightening cosmological constraints from the pessimistic to the optimistic scenario (see Table~\ref{TAB:ERO_ANALYSIS}) and the separate contributions from cluster counts and spatial clustering. In the bottom section, we focus on the optimistic scenario, from {\it eRASS}:8 alone or in combination with other leading probes. See Section \ref{SEC:FISHER} for the operational definition of FoM that is adopted throughout: note that all the FoM values quoted by the DETF refer to forecasts of current or future probes in combination with {\it Planck} priors.}
\EC
\end{figure*}

Differently from the case where also primordial non-Gaussianity of the local type is included \citep[as in ][]{Pillepich:2012}, the constraints in Fig.~\ref{FIG:FM_MAINRESULTS} are dominated by the abundance of the clusters rather than by their clustering signal contribution: see Fig.~\ref{FIG:FM_COMP}, upper panel. Specifically, the combination of counts and clustering improves upon the sole clustering experiment by a factor of about 4 and 30 in $\Delta \SIGMA8$ and in the DE FoM, respectively, while the improvements wrt to the abundance experiment alone read about 70 per cent and 60 per cent, respectively.

We also find that what mostly drives the tightening of the constraints from the pessimistic to the optimistic scenario is (i) the better knowledge of the LM relation, especially for $\SIGMA8$ and $\OM$, and (ii) the lower cluster-mass threshold, i.e. larger number of objects, particularly for the DE sector. Interestingly, when also group-size objects are included in the analysis and pessimistic priors on the LM sector are adopted, errors on $w_0$ and $w_a$ shrink by about an additional 20-30 per cent in comparison to the case when only high-mass objected are included in the analysis. The resulting improvement on the DE FoM (Fig.~\ref{FIG:FM_COMP}, top panel) is similar to the case where the uncertainties on the LM relation are reduced by a factor of four at the cluster-mass scale (as described in Section~\ref{SEC:SELFCAL}).

Within our framework, results do not seem to depend on the accuracy of the redshift measurements,
once reasonable priors are adopted on the LM relation and at least a few bins in redshift are considered.
This implies that the results of our optimistic scenario will not sensibly deteriorate if spectroscopic redshifts will not be available.
However, they will still be of pivotal importance for point-source identification, for the measurement of redshift-space distortions, and for the identification of (admittedly rare) structures in projection.

\begin{table}
\begin{center}
\caption{\label{TAB:RESULTS_COMPARISON} Comparison among $\eRO$ (optimistic case), {\it Planck} with and without additional data, and the DES results, for cosmological models with constant or evolving DE (see Section \ref{SEC:RESULTS_COMPARISON} for details). The results for the 1st-year constraints from the DES are from \textcolor{blue}{Troxel et al. 2017}. As a future, specific example of StageIV experiment, we give the predictions for the {\it Euclid} mission according to \textcolor{blue}{Giannantonio et al. 2012} (weak-lensing + 2D spectroscopic clustering of galaxies) and \textcolor{blue}{Sartoris et al. 2016} (cluster counts and 3D power spectrum but assuming {\it perfect knowledge} of the cluster mass-observable scaling relations). 
Notes: (1) {\it Planck} = base\_plikHM\_TTTEEE\_lowTEB; {\it Planck} + BAO + JLA = base-w-wa\_plikHM\_TTTEEE\_lowTEB\_BAO\_H070p6\_JLA.
(2) Here, perfect knowledge of the cluster scaling relations is assumed which leads to very optimistic constraints.}
\begin{tabular}{l c c c c }
\hline
&&&&\\ 
Data	& $\Delta\SIGMA8$ & $\Delta\OM$ & $\Delta w_0$ & $\Delta w_a$ \\
&&&&\\
\hline

&&&&\\

{\it Planck}							    & 0.013	    	& 0.009			& - 			& -    \\
{\it Planck} + BAO + JLA					& 0.017	    	& 0.010			& 0.11 			& 0.41 \\
&&&&\\
{\it eRASS}:8 						& 0.015 	& 0.010 		& 0.098 		& 0.39 \\
{\it eRASS}:8 + {\it Planck}$^{(1)}$ 				& 0.008 	& 0.006 		& 0.082 		& 0.33 \\
{\it eRASS}:8 + {\it Planck} + BAO + JLA  	& 0.008 	& 0.006 		& 0.068 		& 0.25 \\

&&&&\\
DES 1st year 						& - 	& $^{+0.074}_{-0.054}$		& $^{+0.26}_{-0.48}$	& - \\
&&&&\\

{\it Euclid} + {\it Planck} (wl+gc) 			& 0.005	&0.004		& 0.035		& 0.15	\\
{\it Euclid} + {\it Planck} (clusters)$^{(2)}$			& 0.001	&0.001		& 0.034		& 0.16	\\

&&&&\\

\hline
\end{tabular}
\end{center}
\end{table}

\subsection{Combined constraints with {\it Planck} and other data}
\label{SEC:RESULTS_COMPARISON}

In Table \ref{TAB:RESULTS} and Fig. \ref{FIG:FM_MAINRESULTS}, 
we combine the results above
with the analysis of the CMB anisotropies performed by
the {\it Planck} collaboration \citep{PlanckXIII:2015}. 
Specifically, we adopt
the 2015 constraints from the all-sky CMB temperature and polarization analysis from {\it Planck} in combination with BAO data and the JLA sample of Type Ia supernovae\footnote{We use the base-w-wa\_plikHM\_TTTEEE\_lowTEB\_BAO\_H070p6\_JLA chain results, here simply denoted as `{\it Planck} + BAO + JLA', according to the nomenclature of the {\it Planck} data release: http://wiki.cosmos.esa.int/planckpla/index.php/Cosmological\_Parameters.}. 

In the optimistic scenario and within the $\Lambda$CDM model, $\eRO$ alone will return comparable constraints on the {\it derived} parameters $\SIGMA8$ and $\OM$ to the combination of CMB temperature and polarization and BAO and SNe data (thin, black curve in the top panel of Fig. \ref{FIG:FM_MAINRESULTS}). In the case of DE models, the optimistic $\eRO$ dataset alone will outperform {\it Planck}+BAO+JLA data. Finally, in the most general model we consider, the combination with {\it Planck} will further reduce $\eRO$ optimistic constraints to marginalized, one-dimensional, 1$\sigma$ errors as good as $\Delta \SIGMA8 = \pm~0.008$, $\Delta \OM = \pm~0.006$, $\Delta w_0 = \pm~0.07$, and $\Delta w_a = \pm~0.25$.

These results correspond to a DE FoM of about 162 (116) in the optimistic (pessimistic) scenario, placing $\eRO$ at the level of the StageIV DE experiments identified by the DETF. As we show in Fig.~\ref{FIG:FM_COMP}, according to the recommendations of \cite{Albrecht:2006}, StageIII and StageIV experiments in combination with {\it Planck}-like data should be able to constrain DE models with a FoM in the range [8, 43] and [27, 645], respectively.
These values bracket their pessimistic and optimistic assumptions on systematics control and survey characteristics, and combine results from all types of DE techniques (BAO, clusters, supernovae, and weak lensing). Thanks to its $10^5$ clusters, the constraints set by $\eRO$ 
will exceed the DETF expectations from cluster data (FoM in the range [6, ~39], both at the StageIII and StageIV levels), even with conservative assumptions on the LM relation and on redshift availability. In fact, {\it eRASS}:8 will provide StageIV-level results also without {\it Planck} priors.

To better put $\eRO$ into context, in Table \ref{TAB:RESULTS_COMPARISON} we contrast our forecasts (optimistic case) with {\it Planck} constraints (alone and in combination with BAO and SN probes) and the recent results from the 1st-year analysis of the Dark Energy Survey \citep{DES1st}. The {\it Planck} only case (from the base\_plikHM\_TTTEEE\_lowTEB chains) is meant to highlight the complementary contribution on the DE sector of $\eRO$ alone when the latter is combined with other cosmological probes, e.g. BAO and Type Ia SNe. As a future example of StageIV experiment, we show some expectations  for the {\it Euclid} mission \citep{Euclid:2011}, a galaxy and galaxy cluster survey. In particular, we compare to the predictions by \cite{Giannantonio:2012} and \cite{Sartoris:2016}, who had studied the constraining power of the weak-lensing and the 2D spectroscopic clustering of galaxies and of the cluster counts and 3D power spectrum, respectively, the latter assuming {\it perfect knowledge} of the cluster mass-observable scaling relations (see \citealt{Amendola:2018} for a recent review on cosmology with {\it Euclid}).


\subsection{From {\it eRASS:1} to {\it eRASS:8}}
\label{SEC:ERASSn}

So far, we have provided constraints that $\eRO$ will be able to achieve at its final survey stage, namely after eight successive scans of the entire sky. These will be available after four years of observations, i.e. sometime in 2023.
It is, however, interesting to investigate what we can expect from early and intermediate data releases.

\begin{figure*}
\BC
\includegraphics[width=17cm]{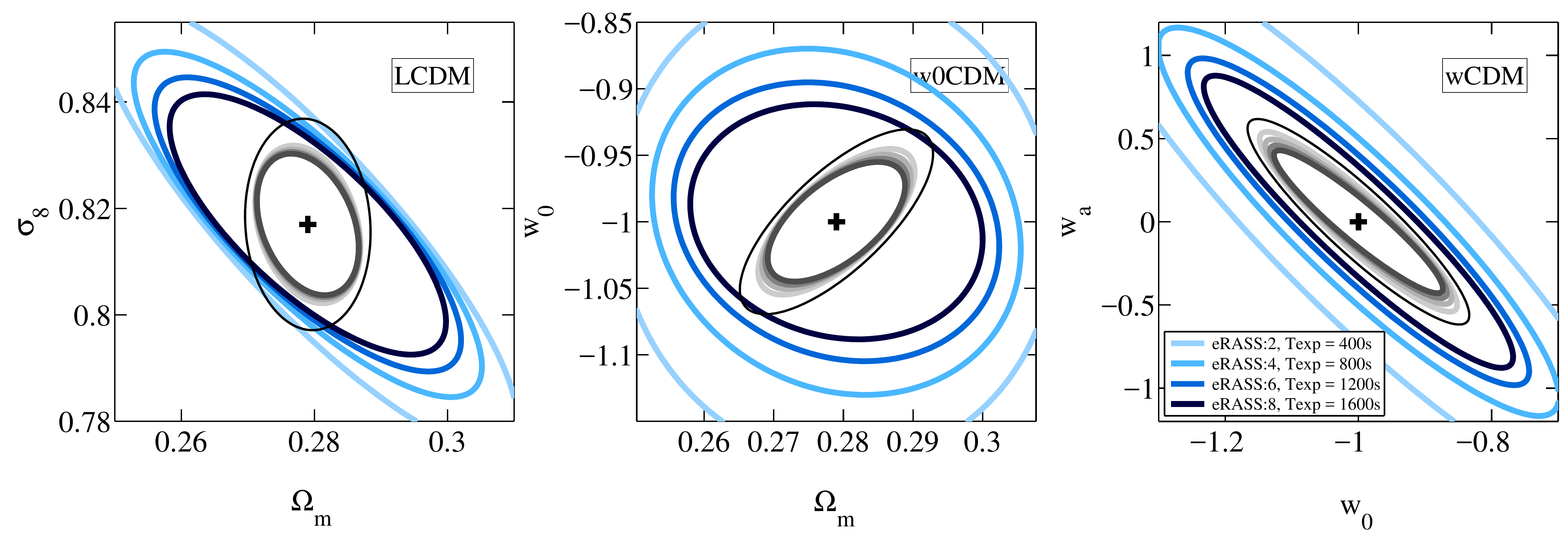}
\includegraphics[width=17cm]{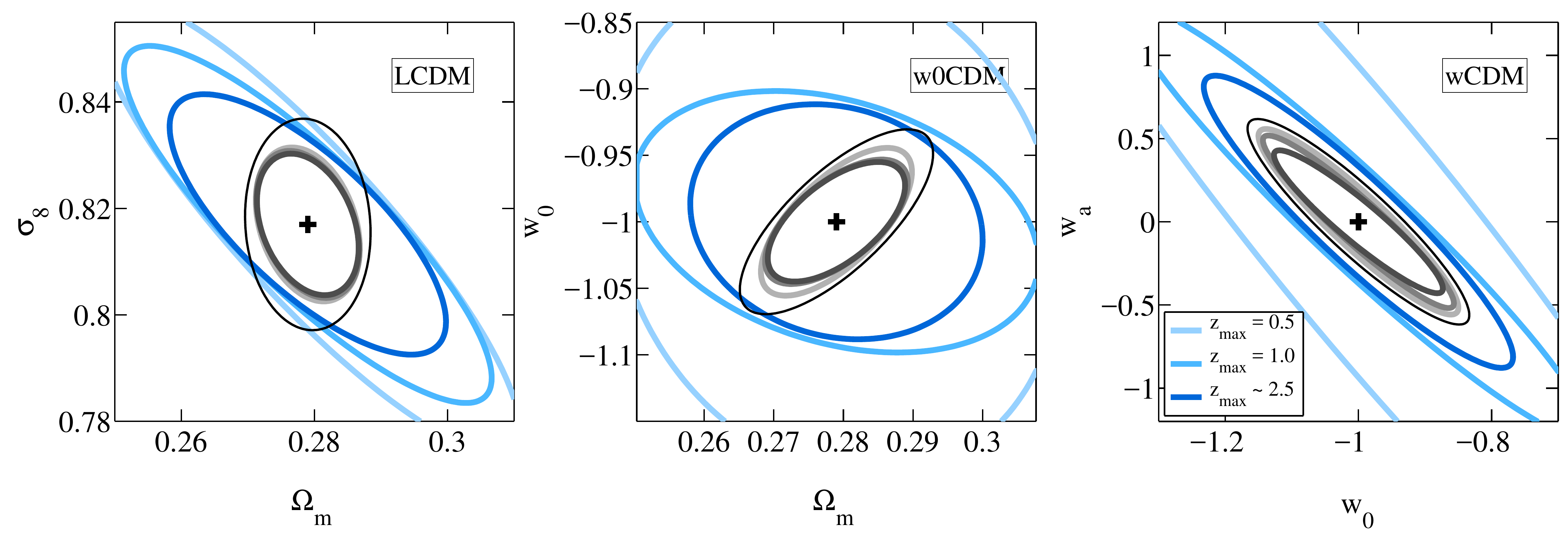}
\caption{\label{FIG:RESULTS_SURVEYS} Top: $\eRO$ constraining power as a function of time i.e. at subsequent stages of its all-sky survey, shown here at annual intervals. Bottom: $\eRO$ constraining power as a function of maximum available cluster redshift. Results are given adopting the pessimistic assumptions described in Table \ref{TAB:ERO_ANALYSIS}, standard priors on $h$ and $\OB$, and without and in combination with {\it Planck} CMB data: blue and gray curve sets, respectively. Black thin contours denote results from the combination of {\it Planck}, BAO and supernova data, for reference.}
\EC
\end{figure*}

In Fig.~\ref{FIG:RESULTS_SURVEYS}, the progression of the $\eRO$ constraints during the four years of its all-sky survey is obtained assuming gradually deeper average exposures. 
Here and in Table \ref{TAB:RESULTS_SURVEYS}, where we provide results for {\it eRASS}:2 and compare them to the final ones, we only consider our pessimistic set-up.
Given that the data will be equally split between the German and the Russian consortia
and considering the unavailability of follow-up campaigns covering the whole sky
before late 2019, in Table \ref{TAB:RESULTS_SURVEYS} we also list constraints obtained with only half-sky coverage. 

Within the $\Lambda$CDM model, $\eRO$ will return per-cent level statistical constraints on $\SIGMA8$ and $\OM$ from its very first year of observations (Fig.~\ref{FIG:RESULTS_SURVEYS}, top left; all-sky case). For a DE model with constant equation of state, {\it eRASS}:2 will be able to provide constraints on $w_0$ better than 15 or 4 per cent alone or in combination with {\it Planck} data, respectively (for both sky coverages, not tabulated). For evolving DE models, $\eRO$ will qualify as a StageIV experiment already after two years of observations (Table~\ref{TAB:RESULTS_SURVEYS}, top sections) and {\it eRASS}:8 will be able to place competitive constraints on DE ($\Delta w_0 = \pm 0.09$, $\Delta w_a = \pm 0.31$) even in the case where data from only half the sky will be included in the analysis (Table~\ref{TAB:RESULTS_SURVEYS}, fourth and eight row, without and with additional data).

The promising outcome of {\it eRASS}:2 displayed in the top panel of Fig.~\ref{FIG:RESULTS_SURVEYS} in combination with {\it Planck} data and even adopting pessimistic assumptions should not be mistaken as a reason to stop the all-sky survey after one year only of observations. In fact, the deeper sky scans will be vital to shed new light on the ongoing debate regarding possible tensions between cluster results and {\it Planck} constraints.
 

\subsection{The impact of high-redshift clusters}
\label{SEC:HIGHZ}

Finally, we quantify the impact of high-redshift clusters on our
forecast. Although the median redshift of the $\eRO$ clusters is $z\sim 0.35$, we expect to find a fourth of the sample (with $\M500c\gtrsim 5 \times 10^{13} h^{-1} \MSUN$)  at $z\geq0.5$ and about 1,100 clusters with $z\geq1$.
As reported in Table \ref{TAB:RESULTS_SURVEYS} for $w$CDM models, while the bulk of the information for $\SIGMA8$ and $\OM$ gets saturated with the low-redshift clusters (especially at $z\lesssim1$), the DE sector largely benefits from the leverage of the objects at higher redshifts. For example, within the pessimistic assumptions and for $\eRO$ alone, the constraints on the EOS parameters would degrade by a factor of $\sim 3$ if we were in the unfortunate conditions to exclude objects above redshift 0.5, because of identification issues, lack of redshift confirmation or poor knowledge of the scaling relations. Even more interestingly, the exclusion of the thousand clusters beyond redshift 1 would provoke a deterioration of the error bars on $w_0$ and $w_a$ of about 50 per cent. 
This result highlights the vital importance of high-redshift clusters for the DE sector, for which good observable-mass calibrations will be needed.


\begin{table*}
\caption{\label{TAB:RESULTS_SURVEYS} $\eRO$ constraining power on the $w$CDM models for different survey strategy configurations. The expectations in the top eight rows correspond to two stages of the all-sky survey: the early case {\it eRASS}:2 (after one year of observations) and the final one {\it eRASS}:8 (after four years), for two different assumptions on the sky coverage, all vs. half sky. Results of {\it eRASS}:2 are in both cases given adopting pessimistic assumptions and are dominated by the {\it Planck} priors; results from {\it eRASS}:8 are obtained with optimistic assumptions in both the all-sky and half-sky cases. On the bottom section of the table, different scenarios for the availability of high-redshift clusters are quantified, with pessimistic assumptions.}
\centering
\begin{tabular}{l l c c c c c c c}
\hline
&&&&&&&\\
Data	 	& $f_{\rm sky} [deg^2]$ & $T_{\rm exp}$ [s] & \# Clusters & $\Delta\SIGMA8$ & $\Delta\OM$ & $\Delta w_0$ & $\Delta w_a$ & FoM$^{\rm DE}$ \\
&&&&&&&\\
\hline

&&&&&&&\\
{\it eRASS}:2 + $H_0$ + BBN 	& 27,145 & 400	& 10,016		& 0.037 		& 0.026 		& 0.28 		& 1.11 	& 7 \\ 
{\it eRASS}:8 + $H_0$ + BBN 	& 27,145 & 1600	& 125,300		& 0.015 		& 0.010 		& 0.10 		& 0.39 	& 69 \\ 
{\it eRASS}:2 + $H_0$ + BBN  	& 13,573 & 400 	& 5,009			& 0.042 		& 0.028 		& 0.37 		& 1.47 	& 5 \\ 
{\it eRASS}:8 + $H_0$ + BBN 	& 13,573 & 1600	& 62,650 		& 0.024			& 0.015 		& 0.20 		& 0.68 	& 26 \\ 

&&&&&&&\\
{\it eRASS}:2 + $H_0$ + BBN + {\it Planck} + BAO + JLA	& 27,145 & 400	& 10,016		& 0.012 		& 0.008 		& 0.10 		& 0.37 	& 74 \\ 
{\it eRASS}:8 + $H_0$ + BBN + {\it Planck} + BAO + JLA	& 27,145 & 1600	& 125,300		& 0.008 		& 0.006 		& 0.07 		& 0.25 	& 162 \\ 
{\it eRASS}:2 + $H_0$ + BBN + {\it Planck} + BAO + JLA 	& 13,573 & 400 	& 5,009			& 0.013 		& 0.008 		& 0.10 		& 0.37 	& 69 \\ 
{\it eRASS}:8 + $H_0$ + BBN + {\it Planck} + BAO + JLA	& 13,573 & 1600	& 62,650 		& 0.010			& 0.007 		& 0.09 		& 0.31 	& 98 \\ 
&&&&&&&\\

$z \lesssim 0.5 $ {\it eRASS}:8 + $H_0$ + BBN		& 27,145 & 1600	& 67,500 		& 0.040		& 0.027 		& 0.39 		& 1.89 	& 4 \\ 
$z \lesssim 1.0 $ {\it eRASS}:8 + $H_0$ + BBN		& 27,145 & 1600	& 87,800 		& 0.030		& 0.022 		& 0.23 		& 0.86 	& 18 \\ 
$z \lesssim 2.5 $ {\it eRASS}:8 + $H_0$ + BBN		& 27,145 & 1600	& 88,900 		& 0.029		& 0.022 		& 0.15 		& 0.58 	& 29 \\ 
&&&&&&&\\
\hline
\end{tabular}
\end{table*}

\section{Discussion}
\label{SEC:MODELS}

The scope of this paper is to provide the community with baseline expectations for the $\eRO$ constraints on DE, by focusing on the abundance and spatial clustering of its photon-count limited, all-sky, galaxy-cluster sample. So far, we have focused on the {\it statistical} errorbars of the parameters of interest. Here, we comment on possible systematic uncertainties.

\subsection{On the fiducial cosmology and observable--mass relation}
Fisher forecasts require assuming an underlying cosmological model.
Here, we have used the cosmological constraints from \citealt{Komatsu:2009} (5-year {\it WMAP} + BAO and SN data). This choice was dictated primarily by
consistency with the X-ray LM scaling relation derived in \citet{Vikhlinin:2009a,Vikhlinin:2009b}. We have tested the effects of changing the fiducial cosmology to the most recent {\it Planck} results. 
In this case, however, we have artificially decreased the cluster luminosities predicted by the \cite{Vikhlinin:2009a} relation by 40 per cent in order to keep the overall number of $\eRO$ clusters unchanged with respect to our reference case. The forecasts for the cosmological parameters are consistent in the two scenarios. 

Another possible source of systematics in our predictions lies in the adopted observable--mass relation(s). We have replicated our predictions by assuming the X-ray scaling relations by \cite{Reichert:2011}. For the $w$CDM scenario but keeping the observable--mass scaling relation parameters frozen (e.g. as in the `LM:fixed' case of Fig.~\ref{FIG_PRIORS}), we recover the same degeneracies among parameters as with the \cite{Vikhlinin:2009a} relations: the discrepancies in the statistical forecasts for the two sets of assumptions are smaller than about 10 per cent, for all cosmological parameters but for the spectral index and the baryonic cosmic fraction. However, for this comparison we have assumed the same LM intrinsic scatter, namely $ \SIGMALM = 0.396$, as per \cite{Vikhlinin:2009a} findings. A larger scatter as the one found by \citealt{Andreon:2016} ($ \SIGMALM = 0.47$) would actually translate into somewhat tighter statistical constraints, by no more than 10 per cent in our parameters of interest in the cluster abundance only calculation. 

\subsection{On the underlying theoretical models}

Another aspect of a Fisher forecast is that 
it does not require extremely accurate models of the observables. 
Basically, the forecast relies on the assumption that 
accurate models will be available when the actual data will be analyzed. It is only in this case that the potential of the survey is fully realized in practice.
However, even in a forecast, it is necessary to make sure that the underlying model grasps the key features, trends, and parameter degeneracies that characterize the survey, as, in our case, the cluster number counts as a function of redshift and $\eta$.

In this respect, we remark that the fitting functions we have used for the halo mass function \citep{Tinker:2008} have not been calibrated for DE models different from a cosmological constant \citep[but see][]{Bhattacharya:2011, Cui:2012a}.
The \citet{Tinker:2008} model is accurate at percent level {\it at fixed cosmology} and shows about 5 per cent deviations within the range of $\Lambda$CDM cosmologies but possibly worse accuracies for evolving-DE cosmological models.
Moreover, additional uncertainties of a few per cent
are introduced by the numerical interpolation which is needed
to convert the Tinker's formulas 
into functions of $\Delta_{\rm crit}$.
Therefore, we cannot exclude that our results are affected
by uncertainties of this size or slightly larger.

Even more importantly, our approach neglects the effects of baryonic physics \citep[as we use the DM halo mass function and bias by][]{Tinker:2008, Tinker:2010a}. Over the last years, a number of studies have investigated the impact of cluster gas physics on the functional form of the halo mass function via hydrodynamical cosmological simulations \citep{Cui:2012b, Sawala:2013, Cusworth:2014, Vogelsberger:2014b, Bocquet:2016} and thus on the number of expected galaxy clusters and associated cosmological constraints \citep{Martizzi:2012, Balaguera:2013,Bocquet:2016}. AGN feedback, galactic winds, and complex ICM physics have been shown to modify the expectations from dark-matter only simulations on a host of observables, including the DM and total-matter-density profiles within clusters, the meaning of cluster mass, the clustering of dark and luminous matter across spatial scales and times, and the DM halo and galaxy bias \citep[see e.g.][for Illustris, EAGLE, Horizon-AGN, and IllustrisTNG results]{Vogelsberger:2014b, Schaller:2015, Chisari:2018,Springel:2018}. Owing to the complexity of the mechanisms at work, no consensus has yet been reached across simulation campaigns and subgrid physics prescriptions as to how and by how much the N-body fitting halo mass function and halo bias formulas shall be modified to take into account the effects of baryons. Yet, the majority of the researchers
conducting hydrodynamical numerical campaigns agree that such modifications far exceed the accuracies required to fulfill the promises of the era of precision cosmology.

In fact, because of the enormous computational cost that is needed to simulate very massive structures or very large cosmological volumes by simultaneously resolving the details of the intra-cluster plasma, of the cluster galaxies, and hence of the interplay between galaxies and feedback, currently no baryonic halo mass function predictions extending to a few $10^{15} \MSUN$ can be considered statistically sound. We therefore conclude that, while our results are most certainly affected by a non negligible systematic uncertainty, the landscape of baryonic predictions is still not fully mature for the problem at hand and a thorough quantitative assessment has to be postponed to future studies.

\begin{figure*}
\BC
\includegraphics[width=15.45cm]{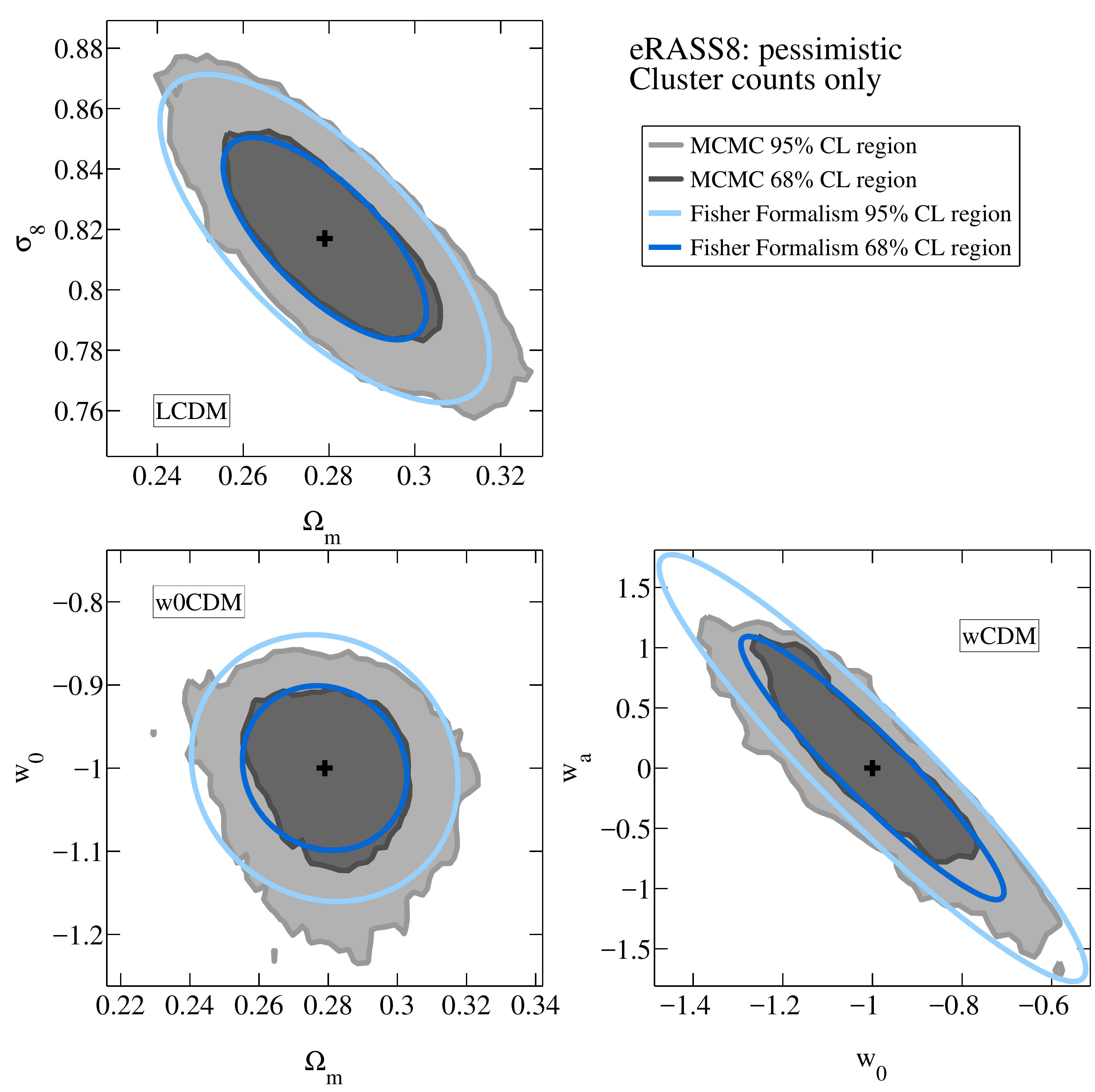}
\caption{\label{FIG:FM_MCMC} Test of the cosmological constraints obtained with the Fisher-matrix method against a full likelihood analysis.
We consider the cluster abundance only for {\it eRASS}:8 in a pessimistic scenario. 
The ellipses denote the 68.3 and 95.4 per cent credible regions obtained with the Fisher formalism and are similar to those reported in Fig. \ref{FIG:FM_MAINRESULTS} -- where, however, also the clustering is considered.
The shaded areas mark the 68.3 and 95.4 per cent credible regions obtained by sampling the corresponding posterior distribution with a MCMC.}
\EC
\end{figure*}

\subsection{Fisher matrix vs sampling the posterior distribution}
The Fisher information matrix is a precious statistical tool for assessing the constraining power
of planned experiments. However, this method has limitations and its results do
not always agree with those obtained computing the posterior distribution based on Bayes theorem.
The main assumption of the Fisher formalism is that the
likelihood function (or the posterior distribution, depending on the application) averaged over all possible datasets is a multi-variate Gaussian. 
This is a strong hypothesis and
examples have been provided showing
sets of parameters
for which the Fisher-matrix constraints do not match 
the actual contour levels of the likelihood function \citep[e.g.][]{Wolz:2012, Khedekar:2013}.
Typically, problems arise for highly degenerate parameters with large
uncertainties. In these cases, it is
recommended to
reparameterize the model using
combinations of the degenerate variables before computing a Fisher forecast  
\citep[e.g.][]{Albrecht:2006}.

In order to validate our Fisher forecasts, in Fig.~\ref{FIG:FM_MCMC} we compare them with
those obtained by sampling the full posterior distribution with a Monte Carlo Markov Chain (MCMC). Specifically, we use the
Metropolis sampler included in the
{\sc cosmoMC} software \citep[][version from 06/2015]{Lewis:2002} together with our own likelihood function for the $\eRO$ clusters.
For this test,
we only consider the cluster number counts that, as shown above, dominate the cosmological constraints.
We find that the marginalized
68 per cent credible intervals
are consistent to better than 17 per cent for all the parameters and cosmological models we have considered.
In particular, the marginalized Fisher
errors on $w_0$ and $w_a$ in the evolving
DE scenario are slightly overestimated.
In most cases, the Gaussian approximation for the likelihood function is excellent, especially close to the peak. However, occasionally, mild asymmetries are present
(see e.g. the 95 per cent credible region in the bottom-left panel of 
Fig.~\ref{FIG:FM_MCMC}).
Based on this analysis, we conclude that the Fisher-matrix formalism 
is a reliable forecasting technique
for the purposes of our paper.

\section{Summary and Conclusions}
\label{SEC:CONCLUSIONS}

We have forecast the potential of the upcoming X-ray telescope $\eRO$ to constrain cosmological parameters, with particular attention to DE models. $\eRO$ will be launched in 2019 and will provide the first all-sky, X-ray selected galaxy-cluster sample after {\it ROSAT}, at an average exposure of 1.6 ks in the (0.5-2) keV  energy band. The $\eRO$ catalog will consist of about 89,000 clusters more massive than $5 \times 10^{13} h^{-1} \MSUN$ with a median redshift of $z \sim 0.35$; or of more than 125,000 objects when also groups of galaxies down to a mass limit of $10^{13} h^{-1} \MSUN$ are included (assuming a detection limit of 50 photons).

Our Fisher-matrix analysis is based on the measurement of the abundance and angular clustering of $\eRO$ clusters and considers spatially-flat cosmological models with either constant ($\Lambda$CDM and $w_0$CDM) or time-dependent ($w$CDM) DE equation-of-state parameters.  
Our most general case considers eleven model parameters, seven of which characterize the cosmological model, while the remaining four describe the luminosity--mass relation for X-ray selected clusters.
We find that, after four years of all-sky survey, $\eRO$ will give marginalized, one-dimensional, 1$\sigma$ errors of $\Delta \SIGMA8 = \pm~0.011$, $\Delta \OM = \pm~0.007$, $\Delta w_0 = \pm~0.08$, and $\Delta w_a = \pm~0.29$ in a pessimistic scenario; and of $\Delta \SIGMA8 = \pm~0.008$, $\Delta \OM = \pm~0.006$, $\Delta w_0 = \pm~0.07$, and $\Delta w_a = \pm~0.25$ in an optimistic scenario, in both cases in combination with and improving upon currently existing {\it Planck}, BAO and SNIa data (see Table \ref{TAB:RESULTS} and Fig.~\ref{FIG:FM_MAINRESULTS}).

For the sets of parameters considered here, the constraining power of the $\eRO$ observations is dominated by the cluster number counts (see Fig.~\ref{FIG:FM_COMP}).
In fact, shot noise and strong degeneracies among different parameters limit the information that can be retrieved from spatial-clustering studies. 
We also find that what mostly drives the tightening of the constraints from the pessimistic to the optimistic scenarios is the more accurate knowledge of the luminosity--mass relation (particularly for the $\SIGMA8$ and $\OM$ sector) {\it and} the possibility of extending it to galaxy groups with lower masses while keeping the same accuracy (particularly for the DE sector, see Figs.~\ref{FIG_PRIORS} and \ref{FIG:FM_COMP}). 
With pessimistic (i.e. current) knowledge of the luminosity--mass relation for massive clusters, the leverage on the DE sector of high-redshift $\eRO$ clusters will be vital, with constraints on $w_0$ and $w_a$ degrading by about a factor of 1.5 (3) when clusters above $z\sim1$ ($z\sim0.5$) are excluded from the analysis or cannot be reliably placed in redshift space. This result highlights the importance of securing good mass calibrations also at high redshift. 

The forecasts presented in this work agree to better than 20 per cent 
with those obtained from a full study of the likelihood function
using an MCMC sampler (see Fig.~\ref{FIG:FM_MCMC}). This shows that the Fisher-matrix formalism provides a reliable tool for the problem at hand. 

We can therefore conclude that, with an expected DE FoM ranging between 116 and 162, $\eRO$ will be one of the first StageIV experiments to come online according to the classification of the Dark Energy Task Force. It will improve upon current error bars on $\SIGMA8$ and $\OM$ after its very first year of observations, at which point the DE equation-of-state parameter $w_0$ (in the $w_0$CDM model) will be known to better than 14 per cent (4 per cent) without (with) including {\it Planck} data, even if only half sky is considered (see Fig.~\ref{FIG:RESULTS_SURVEYS}, top panels, and Table~\ref{TAB:RESULTS_SURVEYS}).

However, our findings call for ambitious, synergistic follow-up programs in order to (i) reduce uncertainties in the X-ray luminosity--mass relations by at least a factor of four wrt to current constraints; (ii) extend their characterizations at similar levels of accuracy down to halo masses of $10^{13} \HI\MSUN$; and (iii) equip $\eRO$ clusters with redshift estimates characterized by statistical errors below 0.05(1+$z$) also at redshifts larger than $z\gtrsim 0.5$. Simultaneously, these results call for a similarly ambitious and robust program of numerical calculations aimed at providing statistically-sound theoretical models with which to fit the forthcoming data. Such models will have to take into account a wide range of physical processes that may affect the distribution of (dark) matter within clusters, in order to improve upon (or confirm) gravity-only predictions.

\section*{Acknowledgements}
The authors would like to thank Irshad Mohammed for the many calculation comparisons made over the past years, Nicolas Clerc and Tim Schrabback for comments on a draft version of this paper, and the $\eRO$ collaboration in general. 
THR and CP acknowledge support from the German Research Association (DFG) through the Transregional Collaborative Research Centre TRR33 `The Dark Universe'. 
THR also acknowledges support by the German Aerospace Agency (DLR) with funds from the Ministry of Economy and Technology (BMWi) through grant 50 OR 1514. 
This research has made use of NASA's Astrophysics Data System, the {\sl CAMB} code, and the {\sc cosmoMC}  package. Figures and codes in this paper were constructed with Matlab (2013b). 

\bibliographystyle{mnras}
\bibliography{Pillepich2014_Bibliography}

\end{document}